# The First Cold Antihydrogen[*]

M.C. Fujiwara[ab†], M. Amoretti[c], C. Amsler[d], G. Bonomi[e], A. Bouchta[e], P.D. Bowe[f], C. Carraro[cg] C.L. Cesar[h], M. Charlton[i], M. Doser[e], V. Filippini[j], A. Fontana[jk], R. Funakoshi[b], P. Genova[j], J.S. Hangst[f], R.S. Hayano[b], L.V. Jørgensen[i], V. Lagomarsino[cg], R. Landua[e], D. Lindelöf[d], E. Lodi Rizzini[l], M. Macri[c], N. Madsen[f], M. Marchesotti[j], P. Montagna[jk], H. Pruys[d], C. Regenfus[d], P. Rielder[e], A. Rotondi[jk], G. Testera[c], A. Variola[c], D.P. van der Werf[i]

(ATHENA Collaboration)

[a] Department of Physics, University of Tokyo, Tokyo 113-0033 Japan
[b] Atomic Physics Laboratory, RIKEN, Saitama, 351-0189 Japan
[c] Istituto Nazionale di Fisica Nucleare, Sezione di Genova, 16146 Genova, Italy
[d] Physik-Institut, Zürich University, CH-8057 Zürich, Switzerland
[e] EP Division, CERN, Geneva 23 Switzerland
[f] Department of Physics and Astronomy, University of Aarhus, DK-8000 Aarhus, Denmark
[g] Dipartimento di Fisica, Università di Genova, 16146 Genova, Italy
[h] Instituto de Fisica, Universidade Federal do Rio de Janeiro, Rio de Janeiro 21945-970, Brazil
[i] Department of Physics, University of Wales Swansea, Swansea SA2 8PP, UK
[j] Istituto Nazionale di Fisica Nucleare, Sezione di Pavia, 27100 Pavia, Italy
[k] Dipartimento di Fisica Nucleare e Teorica, Università di Pavia, 27100 Pavia, Italy
[l] Dipartimento di Chimica e Fisica per l'Ingegneria e per i Materiali, Universit`a di Brescia, 25123 Brescia, Italy

**Abstract**

Antihydrogen, the atomic bound state of an antiproton and a positron, was produced at low energy for the first time by the ATHENA experiment, marking an important first step for precision studies of atomic antimatter. This paper describes the first production and some subsequent developments.

## 1. Introduction

One of the ultimate goals of antihydrogen studies is to hold the antiatom in a stable neutral trap to facilitate interrogation of its detailed properties. An important step towards this goal was achieved recently by ATHENA's first production of cold antihydrogen atoms[1].

---

The observations of a small number of relativistic antihydrogen atoms were reported several years ago at CERN[2] and at FERMILAB[3]. However, only atoms with very low velocity can be studied with precision and trapped in a neutral atom trap, hence intense efforts were devoted to produce cold antihydrogen by two competing experiments at CERN's Antiproton Decelerator (AD) facility. A few months after ATHENA's success ATRAP reported corroborating results on the production of cold antihydrogen[4], but with a very different detection technique.

Fundamental motivations for cold antihydrogen studies are at least two-fold. First, via a precision test of CPT invariance, physics beyond the Standard Model (or any standard quantum field theories) will be investigated, possibly probing Planck scale physics. Impressive progress in precision studies of ordinary hydrogen atoms[5,6] will be exploited by their direct comparisons with antihydrogen. Second, a measurement of Earth's gravitational acceleration on antimatter will test non-Newtonian gravitational interactions. Cosmological baryon asymmetry presents yet another motivation for studying antimatter in the laboratory. See Ref. [7] for further discussions.

**2. The Cold Antihydrogen Challenges**

It may seem trivial to simply combine the two charged (anti)particles that attract each other, but it took many years of development to achieve the production of cold antihydrogen atoms[7]. The main challenge was trapping and cooling of large numbers of antiparticles, which are produced at high energies. The antiprotons at the AD are produced from 26 GeV/c protons colliding with a copper target, and collected at 3.5 GeV/c. Deceleration and cooling by some 12 orders of magnitude is required to obtain meV antiprotons. Pioneering contributions by the TRAP collaboration at LEAR is noteworthy in this respect[8]. As for positrons, obtained from radioactive decay of $^{22}$Na source, we need some 9 orders of magnitude reduction in the energy.

Energy-momentum conservation requires that a third particle needs to be involved in order for antihydrogen to be formed. The two main processes responsible for antihydrogen production are thought to be spontaneous radiative recombination and three-body recombination[9]. In the first case, the energy is carried away by a (real) photon, and in the second case, by a neighboring positron (via a virtual photon). The antihydrogen production rate per antiproton for these processes is proportional to $n_{e+}$ and $n_{e+}^2$, respectively, where $n_{e+}$ is the density of positrons overlapping the antiproton cloud. There is possibly also a combination of these two, known as collisonal-radiative recombination. The number of antiprotons that can be captured is determined by the AD parameters, hence the production rate depends crucially on the number and the density of the positrons. ATHENA's powerful positron accumulator[10] is one of the essential components for the production of large numbers of antihydrogen atoms.

Production of cold antihydrogen can only be established by detection of its unambiguous signatures. The ATHENA antihydrogen detector[11] was designed to do just this by simultaneous detection of antiproton and positron annihilations at the same place. High granularity of both the charged particles and gamma-ray detection systems was essential in discriminating the antihydrogen signal against background events.

The design choice of the ATHENA features an open and modular philosophy (Fig. 1). While sealed vacuum is useful for storing antiprotons for months[8], it introduces some complications in the introduction of positrons and laser lights. ATHENA's approach instead allows the powerful positron trapping and accumulation method based on the buffer gas method[10,12], while maintaining the vacuum at a level where antiproton annihilations with residual gas are negligible. The high number and density of positrons, as well as open access to the production region, will be become even more important for the next phases of antihydrogen research, namely when performing laser spectroscopy.

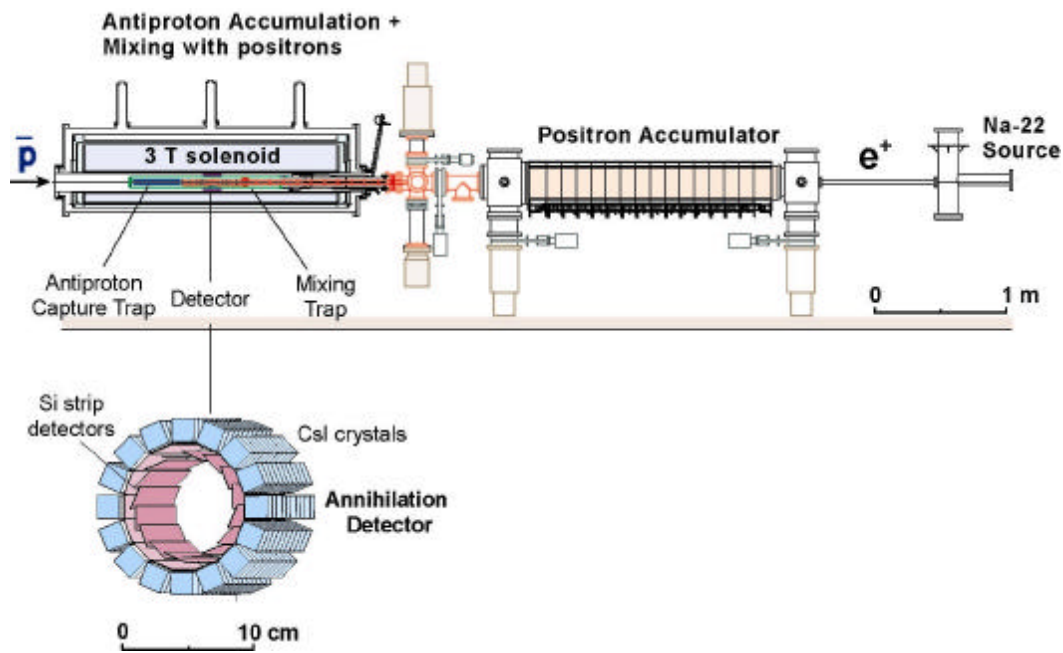

**Figure 1. An overview of the ATHENA antihydrogen apparatus.**

## 3. The Production and Detection

The AD provides a bunch of a few times $10^7$ antiprotons having a kinetic energy of 5 MeV every 100 s. These antiprotons are slowed by degrader foils, and captured dynamically using a pulsed electric field. Typically $10^4$ antiprotons are trapped in the antiproton capture trap with a 5 x $10^{-4}$ trapping efficiency. Electrons, preloaded in the catching trap, cool the antiprotons via Coulomb collisions[8,13]. The electrons in turn are cooled by emission of synchrotron radiation in a 3 T magnetic field. The cooled antiprotons are then transferred to the mixing trap. Figure 2 shows the procedure for antiproton catching in a Penning trap, where particle's radial motion is confined by a uniform magnetic field and the axial motion by an electric field.

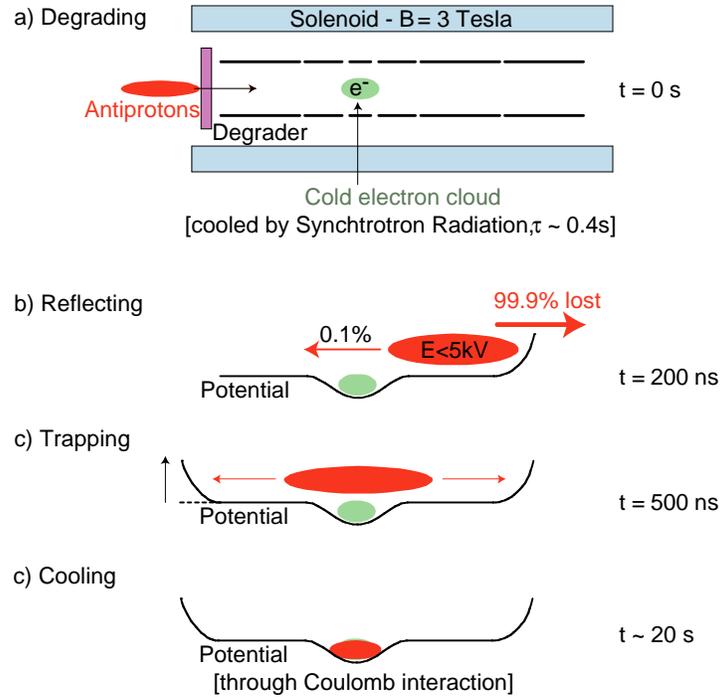

**Figure 2:** Schematic illustration of the procedures of antiproton trapping and cooling in a Penning trap.

Relativistic positrons from a 40 mCi $^{22}$Na source are moderated via a frozen neon moderator held at 5 K, and accelerated in vacuum with a 50 V bias potential. Guided by magnetic coils, the positrons are injected into the accumulator trap, in which they collide inelastically with a nitrogen buffer gas, losing kinetic energy. About 1.5 x $10^8$ positrons are thus accumulated every 5 minutes, a rate three orders of magnitude larger than an alternative scheme[14] when normalized to the source activity. The accumulated positrons are injected into the main magnet and re-trapped in the mixing trap with a 50% efficiency.

The antiprotons and positrons are mixed in the mixing trap, where a double wall potential configuration, so called "nested trap", allows simultaneous confinement of opposite charged particles (Fig. 3). The trap configuration is similar to the one proposed in Ref. [15], but the details of the potential structure turn out to be important in avoiding particle losses, and in allowing efficient mixing. For our first report[1], $10^4$ antiprotons were mixed with 7.5 × $10^7$ positrons for 190 seconds, and this cycle was repeated 165 times.

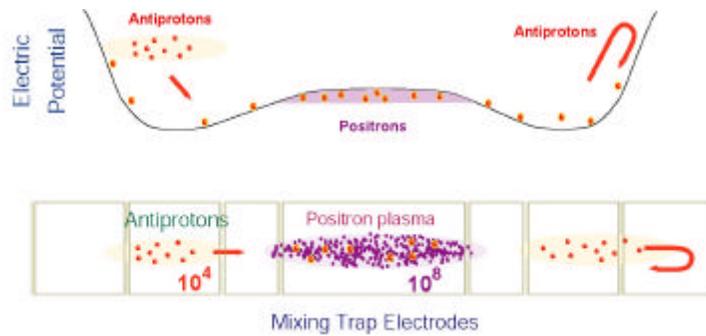

**Figure 3: A schematic of antiproton-positron mixing in a double wall trap.**

If antihydrogen atoms are formed, they escape the electro-magnetic confinement of the mixing trap, and drift to the wall where they annihilate (Fig 4). Antiprotons annihilate to charged pions which are detected by two layers of double-sided silicon microstrip detectors, while positron annihilation produces predominantly two 511 keV gammas which are detected by highly segmented CsI crystals read out via avalanche photodiodes.

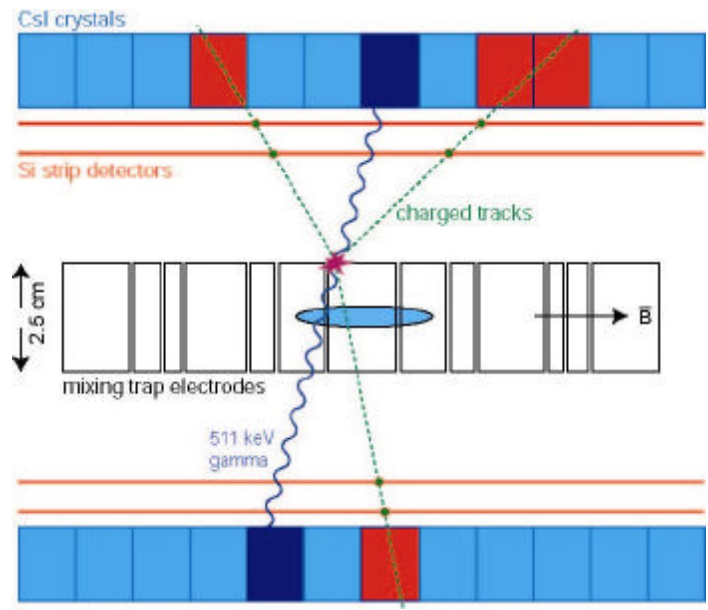

**Figure 4. A schematic view of the mixing trap and the antihydrogen detector**

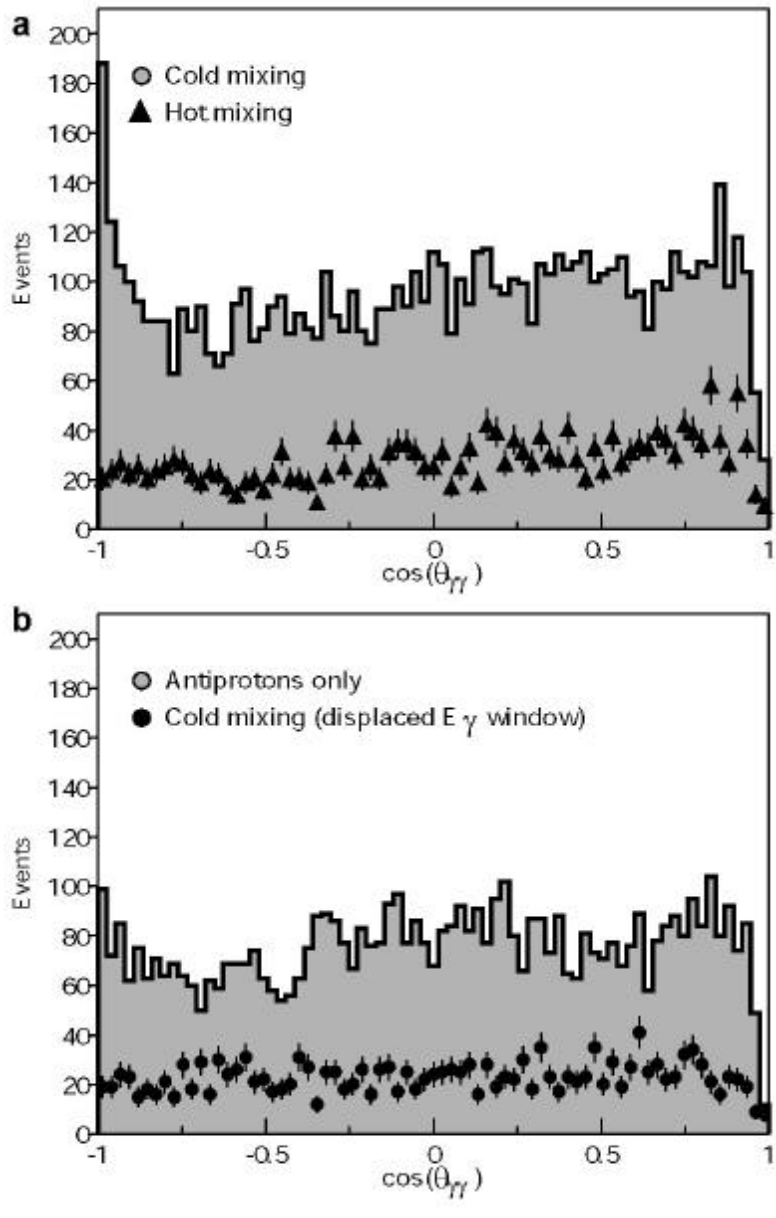

**Figure 5:** The angle between two 511 keV gamma rays, as seen from the reconstructed vertices of antiproton annihilations **(a):** antiproton mixing with cold positrons (grey histogram), and that with positrons heated with RF to above 3000 K (triangle). **(b):** without positrons, and only antiprotons annihilating on the electrode wall (grey histogram), and standard cold mixing data but analyzed with a gamma energy window displaced to a region above 511 keV (circle).

In Fig. 5 (a: histogram), we plot the opening angle of two 511 keV gamma rays, seen from the antiproton annihilation vertex. Gamma events are filtered by demanding off-line that there be no hits in the neighboring eight crystals, and that no charged tracks pass through the crystal. The peak at $\cos(\theta_{\gamma\gamma}) \approx -1$, thus obtained, represents the events in which the positron and the antiprotons annihilate in proximity at the same time, and is indicative of antihydrogen annihilation at the trap wall. The prime source of the background is annihilations of antiprotons themselves; antiproton annihilations create neutral pions (in addition to the charged), which decays into high energy gammas. These gamma rays and electromagnetic showers produced by them in the apparatus structure, can mimic 511 keV gamma events. In addition, annihilations of positrons created in the shower can give real 511 keV gammas. These antiproton induced background events are potentially dangerous, since they give a coincident signal of the charged vertex and the gamma-rays. However, these gamma events originate randomly in the apparatus structure, and are generally not correlated in the opening angle seen by the charged vertex. This background was determined experimentally by letting antiprotons annihilate on the wall in the measurement where there were no positrons present. As shown in Fig. 5 (b: histogram), it does not produce a peak at $\cos(\theta_{\gamma\gamma}) \approx -1$, as expected. All the distributions shown in Fig. 5 are in a good agreement with our Monte Carlo simulations.

A separate background measurement was performed by heating the positron plasma to above 3000 K. This was achieved[16] by applying an RF wave that is resonant with electrostatic plasma modes, known as the Dubin modes[17]. As antihydrogen formation rate is proportional to a negative power of the positron temperature, heating the positron plasma effectively suppresses formation without much affecting other conditions, providing a useful background measurement. Plotted in Fig. 5 (a: triangles) is such a measurement, and no peak at $\cos(\theta_{\gamma\gamma}) \approx -1$ is observed. A further consistency check was obtained by displacing the gamma energy window to above the 511 keV region, and performing the same analysis. As shown in Fig. 5 (b: filled circles), again, there is no peak at $\cos(\theta_{\gamma\gamma}) \approx -1$. In addition, the spatial pattern of the antiproton annihilation vertices during the mixing is consistent with neutral antihydrogen annihilating at the trap wall, providing further confirmation.

Based on this evidence, we reported the first production of cold antihydrogen atoms in August 2002. For our first report, we adopted very restrictive event selection criteria, and furthermore, assumed a conservative background level, which resulted in 131±22 fully reconstructed events ("golden events"). Taking into account the estimated detection efficiency of 0.25%, we reported the first production of cold antihydrogen atoms with a lower limit of 50000. In the meantime, further detailed analysis and comparisons with full Monte Carlo simulations have lead us to a preliminary results that we have produced about one million antihydrogen atoms in the year 2002. Details of the new analyses are the subject of our forthcoming publications.

**4. Slow Antiprotons**

The versatility of the ATHENA apparatus allows some unique measurements, in addition to antihydrogen production discussed above. Here we show one such possibility.

There is considerable worldwide interest in producing slow beam of elementary particles. An intense source of slow muons, in particular, is of interest for stopped muon experiments and perhaps for future neutrino factories or muon colliders[18]. While slow positive muons can be obtained for example from frozen gas moderation[19] (like slow positrons) or via ionization of muonium atoms ($\mu^+e^-$)[20], slow negative muons are much harder to obtain. Antiprotons, with their infinite intrinsic lifetime, could provide a useful substitute for studying negative muon slowing processes. Here we use the ATHENA trap as a time-of-flight analyzer for slow antiprotons emerging from degrader foils.

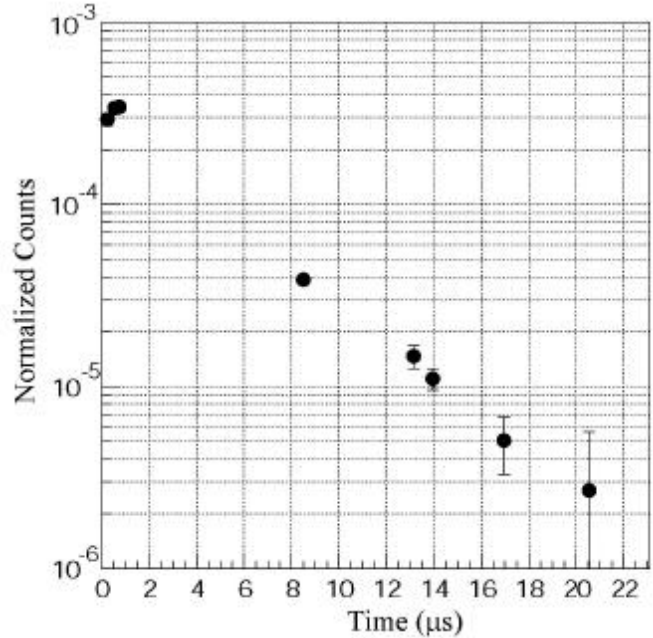

**Figure 6:** The number of captured antiprotons, normalized to antiproton beam intensity, as a function of the time delay between the beam arrival and the closing of the trap potential. See text for details.

As discussed earlier, antiproton capture takes place by opening momentarily the trap potential well at the entrance upon the arrival of the antiproton bunch (Fig. 2). To achieve maximal trapping, the trap potential needs to be closed before the antiproton bunch bounces back to the entrance after traversing a flight path of 2$L$, where $L$=35 cm is the trap length (step C in Fig. 2). By intentionally delaying the trap closing time, we can selectively capture those antiprotons having lower longitudinal energies. Preliminary results shown in Fig. 4 indicates that some antiprotons can be captured even after a ~20 μs delay in closing the trap well, with a trapping efficiency of order $10^{-6}$ with respect to the incoming 5 MeV antiproton beam. Here, the number of the trapped antiprotons was counted by scintillators[13], while the beam intensity was measured by external beam counters read out by hybrid photodiodes[21]. The observed time of flight of 20 μs corresponds to an antiproton longitudinal energy of ~6 eV. If slow muons can be obtained in a similar manner, it would open up possibilities for some future experiments such as production of anti-muonium ($\mu^- e^+$) in a Penning trap. Note, however, that the perpendicular component of the energy could be substantially larger. It should also be mentioned that time-of-flight measurements of this kind are more difficult to do with muons, since most muons, due to their 2 μs lifetime, would decay after several μs of flight time.

## 5. Summary and outlook

In this paper, we discussed how the first atoms of cold antihydrogen were produced by the ATHENA experiment. Its open and modular design, which allowed the use of a powerful positron trapping method, together with a unique position sensitive detection capability, was emphasized. Further detailed studies indicate that we have produced about one million cold antihydrogen atoms in year 2002. An example of parasitic measurements in the ATHENA apparatus was also given.

In the near future, laser spectroscopy of antihydrogen atoms will be attempted, first "in beam", i.e. without trapping the neutral anti-atoms. These will be very difficult measurements due to the low expected rates, but we look forward to the challenges ahead.

**Acknowledgements**

We thank J. Rochet, S. Bricola, H. Higaki, and Y. Yamazaki for their valuable contributions, CERN's PS and AD crew for providing the excellent antiproton beam, and M. Hirsh for a critical reading of the manuscript. We gratefully acknowledge the financial support from INFN (Italy), FAPERJ (Brasil), MEXT (Japan), SNF (Switzerland), NSRC (Demark), EPSRC (UK), and the European Union.